\newif\ifdraft
\draftfalse

\newif\ifextended
\extendedtrue

\documentclass[camera,letterpaper,nomarginnotes,nonarrowgutter]{jpaper}

\usepackage[bookmarks=true,breaklinks=true,colorlinks,linkcolor=black,citecolor=blue,urlcolor=black]{hyperref}
\usepackage{listings}
\usepackage{fancyhdr}

\usepackage{datetime}

\usepackage[compress]{cite}
\usepackage{amsmath,amssymb,amsfonts}
\usepackage{algorithmic}
\usepackage{graphicx}
\usepackage{textcomp}
\usepackage{xcolor}
\usepackage{tikz}
\usetikzlibrary{calc}
\usetikzlibrary{fit}

\usepackage[utf8]{inputenc}
\usepackage[T1]{fontenc}

\usepackage{booktabs} 
\usepackage{setspace}
\usepackage[italic]{mathastext}
\usepackage{array}
\usepackage{titlesec}
\usepackage[normalem]{ulem}
\usepackage{multirow}
\usepackage{multicol}
\usepackage{color}
\usepackage[font={small}]{caption}
\usepackage{float}

\usepackage{balance}

\usepackage[acronym,nonumberlist,nowarn]{glossaries}
    \glsdisablehyper
     \newacronym{ADI}{ADI}{Alternating Direction Implicit}
\newacronym{AGU}{AGU}{Address Generation Unit}
\newacronym[longplural={Access History Tables}]{AHT}{AHT}{Access History Table}
\newacronym{AHT-R}{AHT-R}{Access History Table - Read}
\newacronym{AHT-W}{AHT-W}{Access History Table - Write-Back}
\newacronym{AIP}{AIP}{Access Interval Predictor}
\newacronym{AL}{AL}{Added Latency for column accesses}
\newacronym{ASIC}{ASIC}{Application Specific Integrated Circuit}
\newacronym{AVX}{AVX}{Advanced Vector Extensions}
\newacronym[longplural={Basic Block Vectors}]{BBV}{BBV}{Basic Block Vector}
\newacronym{BHT}{BHT}{Branch History Table}
\newacronym{HBM}{HBM}{Hybrid Bandwidth Memory}
\newacronym{DDR4}{DDR4}{Double-Data Rate 4}
\newacronym{MIMD}{MIMD}{multiple-instruction multiple-data}
\newacronym{SIMT}{SIMT}{single-instruction multiple-thread}
\newacronym{SLP}{SALP}{subarray-level parallelism}

\newacronym{NN}{NN}{Neural Network}
\newacronym{BTB}{BTB}{Branch Target Buffer}
\newacronym{CAS}{CAS}{Column Address Strobe}
\newacronym{AMAT}{AMAT}{Average Memory Access Time}
\newacronym{CCD}{CCD}{Column to Column Delay}
\newacronym{AI}{AI}{Arithmetic Intensity}
\newacronym[longplural={Columnar Database Systems}]{CDBS}{CDBS}{Columnar Database System}
\newacronym[longplural={Chip Multiprocessors}]{CMP}{CMP}{Chip Multiprocessor}
\newacronym{CMT}{CMT}{Chip Multithreading}
\newacronym[longplural={Coarse-Grain Reconfigurable Arrays}]{CGRA}{CGRA}{Coarse-Grain Reconfigurable Array}
\newacronym{C-RAM}{C-RAM}{Computational-RAM}
\newacronym{CWD}{CWD}{Column Write Delay}
\newacronym{DBI}{DBI}{Dynamic Binary Instrumentation}
\newacronym[longplural={Database Management Systems}]{DBMS}{DBMS}{Database Management System}
\newacronym{DDR}{DDR}{Double Data Rate}
\newacronym{DEWP}{DEWP}{Dead Line and Early Write-Back Predictor}
\newacronym{DRAM}{DRAM}{Dynamic Random Access Memory}
\newacronym{DSBP}{DSBP}{Dead Sub-Block Predictor}
\newacronym{DSM}{DSM}{Decomposed Storage Manage}
\newacronym{FAW}{FAW}{Four row Activation Window}
\newacronym{FP}{FP}{Floating-Point}
\newacronym{EDP}{EDP}{Energy-Delay Product}
\newacronym{FCFS}{FCFS}{First-Come First-Serve}
\newacronym{FIFO}{FIFO}{Fist In, First Out}
\newacronym{FSM}{FSM}{Finite State Machine}
\newacronym{FPGA}{FPGA}{Field-Programmable Gate Array}
\newacronym[longplural={Functional Units}]{FU}{FU}{Functional Unit}
\newacronym{GAg}{GAg}{Global Adaptive branch prediction using one Global PHT}
\newacronym{GAs}{GAs}{Global Adaptive branch prediction using per-Set PHT}
\newacronym{GCC}{GCC}{GNU Compiler Collection}
\newacronym{GEMS}{GEMS}{General Execution-driven Multiprocessor Simulator}
\newacronym[longplural={General Purpose Processors}]{GPP}{GPP}{General Purpose Processor}
\newacronym{HMC}{HMC}{Hybrid Memory Cube}
\newacronym{HIVE}{HIVE}{HMC Instruction Vector Extensions}
\newacronym{IATAC}{IATAC}{Inter-Access Time per Access Count}
\newacronym{ILP}{ILP}{Instruction Level Parallelism}
\newacronym{IPC}{IPC}{Instructions per Cycle}
\newacronym{ISA}{ISA}{Instruction Set Architecture}
\newacronym{IRAM}{IRAM}{Intelligent RAM}
\newacronym{KIPS}{KIPS}{Kilo Instructions per Second}
\newacronym{LDS}{LDS}{Linked Data Structure}
\newacronym{LFSR}{LFSR}{Linear Feedback Shift Register}
\newacronym{LFMR}{LFMR}{Last-to-First Miss-Ratio}
\newacronym{MLP}{MLP}{Memory-Level Parallelism}

\newacronym{LLC}{LLC}{last-level cache}
\newacronym{LRU}{LRU}{Least Recently Used}
\newacronym{LSU}{LSU}{Load Store Unit}
\newacronym{LTP}{LTP}{Last-Touch Predictor}
\newacronym{LvP}{LvP}{Live-time Predictor}
\newacronym{LWP}{LWP}{Last Write Predictor}
\newacronym{MARSS}{MARSS}{Micro Architectural and System Simulator}
\newacronym{McPAT}{McPAT}{Multi-core Power, Area, and Timing}
\newacronym{MIPS}{MIPS}{Microprocessor without Interlocked Pipeline Stages}
\newacronym{MOB}{MOB}{Memory Order Buffer}
\newacronym{MMU}{MMU}{memory management unit}

\newacronym{MPKI}{MPKI}{Misses per Kilo-Instruction}
\newacronym{MSHR}{MSHR}{Miss-Status Handling Registers}
\newacronym{NAS}{NAS}{Numerical Aerodynamic Simulation}
\newacronym{NDP}{NDP}{Near-Data Processing}
\newacronym{NMP}{NMP}{Near Memory Processor}
\newacronym{NoC}{NoC}{Network-on-Chip}
\newacronym{NPB}{NPB}{NAS Parallel Benchmark}
\newacronym{NUCA}{NUCA}{Non-Uniform Cache Architecture}
\newacronym{NUMA}{NUMA}{Non-Uniform Memory Access}
\newacronym{OoO}{OoO}{Out-of-Order}
\newacronym{OpenMP}{OpenMP}{Open Multi-Processing}
\newacronym{OS}{OS}{Operating System}
\newacronym{PAg}{PAg}{Per-address Adaptive branch prediction using one Global PHT}
\newacronym{PAs}{PAs}{Per-address Adaptive branch prediction using per-Set PHT}
\newacronym{PC}{PC}{Program Counter}
\newacronym[longplural={Pointer-Chasing Engines}]{PCE}{PCE}{Pointer-Chasing Engine}
\newacronym{PCM}{PCM}{Performance Counter Monitor}
\newacronym{PIM}{PiM}{Processing-in-Memory}
\newacronym[longplural={Pattern History Tables}]{PHT}{PHT}{Pattern History Table}
\newacronym{RAPL}{RAPL}{Running Average Power Limit}
\newacronym{RAS}{RAS}{Row Address Strobe}
\newacronym{RAT}{RAT}{Registers Alias Table}
\newacronym{RC}{RC}{Row Cycle}
\newacronym{RCD}{RCD}{RAS to CAS Delay}
\newacronym[longplural={Row-based Database Systems}]{RDBS}{RDBS}{Row-based Database System}
\newacronym{ROB}{ROB}{Reorder Buffer}
\newacronym{RRD}{RRD}{Row to Row activation Delay}
\newacronym{DNN}{DNN}{Deep Neural Network}
\newacronym{ANN}{ANN}{Artificial Neural Network}

\newacronym{pim}{PiM}{Processing-in-Memory}
\newacronym{pid}{PiD}{Processing-in-DRAM}
\newacronym{pnm}{PnM}{Processing-near-Memory}
\newacronym{pnd}{PnD}{Processing-near-DRAM}
\newacronym{pum}{PuM}{Processing-using-Memory}
\newacronym{pud}{PuD}{Processing-using-DRAM}

\newacronym{RP}{RP}{Row Precharge}
\newacronym{RTL}{RTL}{Register Transfer Level}
\newacronym{RTP}{RTP}{Read To Precharge}
\newacronym{RVU}{RVU}{Reconfigurable Vector Unit}
\newacronym{SDP}{SDP}{Skewed Dead-Block Predictor}
\newacronym{SESC}{SESC}{Superescalar Simulator}
\newacronym{SSE}{SSE}{Streaming SIMD Extensions}
\newacronym{SFP}{SFP}{Spatial Footprint Predictor}
\newacronym{SiNUCA}{SiNUCA}{Simulator of Non-Uniform Cache Architectures}
\newacronym{SMT}{SMT}{Simultaneous Multi-Threading}
\newacronym{SIMD}{SIMD}{single-instruction multiple-data}
\newacronym{SPEC}{SPEC}{Standard Performance Evaluation Corporation}
\newacronym{SoC}{SoC}{System-on-Chip}
\newacronym{SPP}{SPP}{Spatial Pattern Predictor}
\newacronym{SRAM}{SRAM}{Static Random Access Memory}
\newacronym{SSOR}{SSOR}{Symmetric Successive Over-Relaxation}
\newacronym{SSV}{SSV}{Search Set Vector}
\newacronym{TLB}{TLB}{Translation Look-aside Buffer}
\newacronym{TLP}{TLP}{Thread Level Parallelism}
\newacronym[longplural={Through-Silicon Vias}]{TSV}{TSV}{Through-Silicon Via}
\newacronym{VWQ}{VWQ}{Virtual Write Queue}
\newacronym{WTR}{WTR}{Write Recovery time}
\newacronym{WR}{WR}{Write To Read delay time}
\newacronym{TRA}{TRA}{triple row activation}

\definecolor{denim}{rgb}{0.08, 0.38, 0.74}
\definecolor{darkolivegreen}{rgb}{0.33, 0.42, 0.18}
\definecolor{dgreen}{rgb}{0.00, 0.75, 0.00}
\definecolor{darkpink}{rgb}{0.88, 0.28, 0.54}
\definecolor{forestgreen}{rgb}{0.0, 0.27, 0.13}
\definecolor{amber}{rgb}{1.0, 0.49, 0.0}
\definecolor{lightyellow}{rgb}{0.980, 0.956, 0.623}
\definecolor{lightblue}{rgb}{0.980, 0.956, 0.623}
\definecolor{darkamber}{rgb}{0.5, 0.19, 0.0}
\definecolor{dkgreen}{rgb}{0,0.6,0}
\definecolor{gray}{rgb}{0.5,0.5,0.5}
\definecolor{lightmauve}{rgb}{0.68,0.4,0.92}
\definecolor{chocolate}{rgb}{0.48, 0.25, 0.0}
\definecolor{dollarbill}{rgb}{0.52,0.73,0.4}
\definecolor{dkdkgreen}{rgb}{0,0.45,0}
\definecolor{gfored}{rgb}{0.580, 0.050, 0.211}
\definecolor{darkwarmgray}{rgb}{0.15, 0.050, 0.05}
\definecolor{ups-truck}{rgb}{0.53, 0.28, 0.21}

\newcommand{\redt}[1]{\textcolor{black}{#1}}

\ifdraft    
        \usepackage[colorinlistoftodos,prependcaption,textsize=tiny]{todonotes}
    \paperwidth=\dimexpr \paperwidth + 4cm\relax
    \oddsidemargin=\dimexpr\oddsidemargin + 2cm\relax
    \evensidemargin=\dimexpr\evensidemargin + 2cm\relax
    \marginparwidth=\dimexpr \marginparwidth + 2cm\relax
    \newcommand{\iey}[1]{\textcolor{dkdkgreen}{#1}}

    \newcommand{\ieycomment}[1]{\todo[size=\scriptsize, linecolor=dkdkgreen, bordercolor=dkdkgreen, backgroundcolor=white]{\textcolor{dkdkgreen}{\textbf{@Ismail:} #1}}}

    \newcommand{\gfcomment}[1]{\todo[size=\scriptsize, linecolor=gfored, bordercolor=gfored, backgroundcolor=white]{\textcolor{gfored}{\textbf{@GF:} #1}}}

    \newcommand{\atbcomment}[1]{\todo[size=\scriptsize, linecolor=denim, bordercolor=denim, backgroundcolor=white]{\textcolor{denim}{\textbf{@atb:} #1}}}

    \newcommand{\omcomment}[1]{\todo[size=\scriptsize, linecolor=purple, bordercolor=purple, backgroundcolor=white]{\textcolor{purple}{\textbf{@OM:} #1}}}

    \newcommand{\outline}[1]{\textbf{\textcolor{orange}{Onur's Outline:}} #1}

\else
    
    \newcommand{\iey}[1]{#1}
    \newcommand{\ieycomment}[1]{}

    \newcommand{\atbcomment}[1]{}

    \newcommand{\gfcomment}[1]{}

    \newcommand{\omcomment}[1]{}

    \newcommand{\outline}[1]{}

\fi

\DeclareRobustCommand\circledtest[1]{\tikz[baseline=(char.base)]{
            \node[shape=circle,fill,inner sep=0pt] (char) {\small\textcolor{white}{#1}};}}

\usetikzlibrary{calc}

\newcommand{\dingOne}{\circledtest{1}}
\newcommand{\dingTwo}{\circledtest{2}}
\newcommand{\dingThree}{\circledtest{3}}
\newcommand{\dingFour}{\circledtest{4}}
\newcommand{\dingFive}{\circledtest{5}}

\makeatletter
\g@addto@macro{\normalsize}{%
  \setlength{\abovedisplayskip}{0pt plus 1pt minus 1pt}
  \setlength{\belowdisplayskip}{0pt plus 1pt minus 1pt}
  \setlength{\intextsep}{2pt plus 1pt minus 1pt}
  \setlength{\textfloatsep}{3pt plus 1pt minus 1pt}
  \setlength{\dbltextfloatsep}{3pt plus 1pt minus 1pt}
  \setlength{\skip\footins}{0pt plus 1pt minus 1pt}
}

 \titlespacing\section{0pt}{2pt plus 0pt minus 0pt}{1.5pt plus 0pt minus 0pt}
 \makeatother

\def\BibTeX{{\rm B\kern-.05em{\sc i\kern-.025em b}\kern-.08em
    T\kern-.1667em\lower.7ex\hbox{E}\kern-.125emX}}

\def\UrlBreaks{\do\/\do-\/\do.\/\do:}

\expandafter\def\expandafter\UrlBreaks\expandafter{\UrlBreaks
  \do\a\do\b\do\c\do\d\do\e\do\f\do\g\do\h\do\i\do\j
  \do\k\do\l\do\m\do\n\do\o\do\p\do\q\do\r\do\s\do\t
  \do\u\do\v\do\w\do\x\do\y\do\z\do\A\do\B\do\C\do\D
  \do\E\do\F\do\G\do\H\do\I\do\J\do\K\do\L\do\M\do\N
  \do\O\do\P\do\Q\do\R\do\S\do\T\do\U\do\V\do\W\do\X
  \do\Y\do\Z}
  
\newcommand{\squishlist}{
 \begin{list}{$\circ$}
  { \setlength{\itemsep}{0pt}
     \setlength{\parsep}{0pt}
     \setlength{\topsep}{0pt}
     \setlength{\partopsep}{0pt}
     \setlength{\leftmargin}{1em}
     \setlength{\labelwidth}{1em}
     \setlength{\labelsep}{0.5em} } }

\newcommand{\squishsublist}{
\begin{list}{$\rightarrow$}
 { \setlength{\itemsep}{0pt}
    \setlength{\parsep}{0pt}
    \setlength{\topsep}{-10em}
    \setlength{\partopsep}{-3pt}
    \setlength{\leftmargin}{1em}
    \setlength{\labelwidth}{1em}
    \setlength{\labelsep}{0.5em} } }

\newcommand{\squishend}{
  \end{list}  }

\newcommand{\ignore}[1]{}

\makeatletter
\g@addto@macro{\normalsize}{%
 \setlength{\intextsep}{2pt plus 1pt minus 1pt}
 \setlength{\textfloatsep}{3pt plus 1pt minus 1pt}
  \setlength{\dbltextfloatsep}{13pt plus 1pt minus 1pt}
}
 \makeatother

\begin{document}
\setstretch{0.989}

\title{{Memory-Centric Computing:\\Solving Computing’s Memory Problem}\vspace{0.22em}}

\author{
{Onur Mutlu}\quad
{Ataberk Olgun}\quad
{\.I}smail~Emir~Y{\"u}ksel\vspace{0.08em}\\
{SAFARI Research Group}\\
{ETH~Z{\"u}rich}
\vspace{-0.7em}}

\maketitle
    \renewcommand{\headrulewidth}{0pt}
    \fancypagestyle{firstpage}{
        \fancyhead{} 
        \fancyhead[C]{
      } 
    \renewcommand{\footrulewidth}{0pt}
    }
  \thispagestyle{firstpage}
\begin{abstract}

Computing has a huge memory problem. The memory system, consisting of multiple technologies at different levels, is responsible for most of the energy consumption, performance bottlenecks, robustness problems, monetary cost, and hardware real estate of a modern computing system. All this becomes worse as modern and emerging applications become more data-intensive (as we readily witness in e.g., machine learning, genome analysis, graph processing, and data analytics), making the memory system an even larger bottleneck. In this paper, we discuss two major challenges that greatly affect computing system performance and efficiency: 1) memory technology \& capacity scaling (at the lower device and circuit levels) and 2) system and application performance \& energy scaling (at the higher levels of the computing stack). We demonstrate that both types of scaling have become extremely difficult, wasteful, and costly due to the dominant {\em processor-centric} design \& execution paradigm of computers, which treats memory as a dumb and inactive component that cannot perform any computation. We show that moving to a {\em memory-centric} design \& execution paradigm can solve the major challenges, while enabling multiple other potential benefits. In particular, we demonstrate that: 1) memory technology scaling problems (e.g., RowHammer, RowPress, Variable Read Disturbance, data retention, and other issues awaiting to be discovered) can be much more easily and efficiently handled by enabling memory to autonomously manage itself; 2) system and application performance \& energy efficiency can, at the same time, be improved by orders of magnitude by enabling computation capability in memory chips and structures (i.e., processing in memory).  We discuss adoption challenges against enabling memory-centric computing, and describe how we can get there step-by-step via an evolutionary path. 
 
\end{abstract}
\glsresetall

\section{Computing's Memory Problem}
\label{sec:problem}
Memory is a central part of a modern computing system. In the {\em processor-centric} paradigm of computing, memory is treated as an inactive component that only serves the demands (i.e., mainly the data load/store requests but also memory maintenance operations like data refresh) of a processor (e.g., CPU, GPU, FPGA, ASIC), without itself having the ability to manipulate data or even manage itself. This paradigm has unfortunately made memory an even bigger bottleneck because: 1) it leads to huge amounts of data movement across the memory hierarchy~\cite{mutlu2019processing,mutlu2020modern,boroumand2018google,boroumand2021google,ghose.ibmjrd19} to serve the needs of a processor, where computation can only be performed; 2) many levels of caches, complex prefetching mechanisms, complicated out-of-order execution and multithreading machinery are added to the processor, which greatly complicates the system and costs area and power (and in many workloads these resources are not beneficial enough as they require high levels of data locality); 3) the massive bit-level and array-level parallelism inherent in the design of memory, which can enable massively parallel computation, is wasted (i.e., most of memory hardware is idle doing nothing useful for computation during execution of programs). Recent works show that main memory alone is responsible for more than 90\% of the system energy when executing commercial edge neural network models~\cite{boroumand2021google}, more than 62\% of the total system energy is wasted on moving data across the memory hierarchy on commonly-used mobile workloads~\cite{boroumand2018google}, the execution times of many workloads are dominated by waiting for memory~\cite{mutlu2003runahead,kanev_isca2015} (e.g., in all Google data center workloads~\cite{kanev_isca2015}) even in state-of-the-art processors that employ almost all of their hardware real-estate (e.g., more than 90\% of the hardware area of a single node~\cite{mutlu2023-mcc-dac-talk}) to tolerate memory access latencies. On top of all this, the cost of main memory (DRAM) alone surpasses the cost of processors (or any other component) in large server systems~\cite{intel-server-cost} and DRAM is responsible for many failures in a data center~\cite{qiao2023exploring,intel-whitepaper-memory-failure,kokolis2025revisiting,meza2015revisiting,wang2017what,beigi2023systematic,sridharan2015memory,chen2021care,
mezaphd,wu2024removing,schroeder2010large,schroeder2009dram,sridharan2012study,sridharan2013feng,hwang2012cosmic,ferreira2014extra}. With exploding data intensity and data access \& storage demands of modern applications (as we see in e.g., generative artificial intelligence, large machine learning models, genome analysis, and video analytics), memory becomes an even larger performance, energy, robustness, and system scaling bottleneck in processor-centric computing systems~\cite{ghose.ibmjrd19,he2025papi,gu2025pim,mutlu2020modern,gholami2024ai,sevilla2022compute}.

This paper discusses two major memory system challenges that span across the computing stack (devices to applications) and greatly affect computing system performance and efficiency. At the circuit/device levels, memory technology \& capacity scaling are becoming increasingly difficult due to the miniscule technology node sizes, causing robustness problems that can greatly impact cost, capacity scaling, reliability, safety, security, and, in turn, both system performance \& energy efficiency. At the system/application levels, performance \& energy scaling is extremely wasteful and costly today because computation capability cannot be efficiently scaled with memory capacity and bandwidth, due to the huge separation and thin connectivity between memory and computation units in the processor-centric paradigm. The fundamental problem, we argue and demonstrate, is that memory cannot perform any computation (or any autonomous operation) by itself.

Moving to a memory-centric computing (MCC) paradigm can fundamentally solve both major challenges, while providing other benefits (e.g., improve system security, reduce system complexity). Memory-centric design \& execution enables memory components that can perform computation and maintenance operations, thereby unleashing the ability to efficiently scale 1) the technology node \& capacity of a memory component by inherently architecting it to manage its own scaling and robustness problems; 2) computation capability and memory bandwidth proportionally to the memory added to a system, since each memory component comes with computation capability. MCC is best when applied to all resources in a system (e.g., SRAM caches, DRAM main memory, NAND flash SSDs, and magnetic tapes). In this work, we focus on applying it to {\em main memory} (a common bottleneck and the major "low-latency" memory that can house large amounts of data) that uses {\em DRAM}~\cite{dennard1968field} technology (the dominant main memory technology, which is evolving into the future).

\vspace{-1pt}
\section{Memory Scaling}
\vspace{-1pt}

\label{sec:memory_scaling}

DRAM technology scaling, which enables higher capacity and reasonable-energy memories that are needed more than ever, has become greatly more difficult today than it was 12 years ago when I gave an invited talk at IMW 2013~\cite{mutlu2013memory} and argued for a system-level approach to solve the then-already-difficult DRAM scaling challenges. A fundamental issue is that as DRAM technology node size becomes smaller, cells and sensing structures become much less reliable due to reduced charge levels and increased noise levels. For example, data retention capability of a DRAM cell gets worse and noisier~\cite{liu2012raidr,liu2013experimental,khan2014efficacy,qureshi2015avatar,patel2017reach} with smaller cell sizes, necessitating higher refresh rates and in-DRAM error-correcting codes~\cite{kang2014co,mineshphd,patel2017reach,patel2019understanding,patel2020beer,patel2021harp,patel2024rethinking}. A prominent and widespread phenomenon that gets worse with technology scaling and threatens the foundations of robust (i.e., reliable, secure, safe) computing is RowHammer~\cite{kim2014flipping,mutlu2017rowhammer,mutlu2019rowhammer,mutlu2023fundamentally}. RowHammer is a read disturbance mechanism, where repeatedly accessing one DRAM row enough times (before rows get refreshed) causes bitflips in physically nearby rows in real commodity off-the-shelf (COTS) DRAM chips. Our original work from 2012 (published in 2014~\cite{kim2014flipping}) that scientifically demonstrated and rigorously analyzed RowHammer showed that RowHammer bitflips can be induced by user-level programs (with no privilege) on real DRAM-based systems under normal operating conditions. The problem has become much worse since then: DRAM chips of all types (e.g., DDRx, LPDDRx, HBMx) with smaller cell sizes are much more vulnerable to RowHammer~\cite{kim2020revisiting,orosa2021deeper,olgun2023dram,olgun2024read,olgun2025variable}. A RowHammer bitflip happens (at the device level) after only a few thousand row activations in cutting-edge DRAM chips~\cite{kim2020revisiting,olgun2025variable,yaglikci2024svard,tugrul2025understanding}. Many works (e.g.,\redt{~\cite{kim2014flipping, seaborn2015exploiting, van2016drammer, gruss2016rowhammer, razavi2016flip, mutlu2017rowhammer, cojocar2019eccploit, mutlu2019rowhammer, frigo2020trrespass, kwong2020rambleed, deridder2021smash, hassan2021utrr, jattke2022blacksmith, kogler2022half, mutlu2023fundamentally, luo2023rowpress, juffinger2024presshammer, marazzi2024risc, orosa2024spyhammer, kang2024sledgehammer,fournaris2017exploiting,poddebniak2018attacking,tatar2018throwhammer,carre2018openssl,barenghi2018software,zhang2018triggering,bhattacharya2018advanced,google-project-zero,rowhammergithub,pessl2016drama,xiao2016one,bosman2016dedup,bhattacharya2016curious,burleson2016invited,qiao2016new,brasser2017can,jang2017sgx,aga2017good,tatar2018defeating,gruss2018another,lipp2018nethammer,van2018guardion,frigo2018grand,ji2019pinpoint,hong2019terminal,cojocar2020rowhammer,weissman2020jackhammer,zhang2020pthammer,yao2020deephammer,tol2022toward,zhang2022implicit,liu2022generating,cohen2022hammerscope,zheng2022trojvit,fahr2022frodo,tobah2022spechammer,rakin2022deepsteal,aydin2022cyber,mus2022jolt,wang2022research,lefforge2023reverse,fahr2022effects,kaur2022work,cai2022feasibility,li2022cyberradar,roohi2022efficient,staudigl2022neurohammer,yang2022socially,islam2022signature,baek2025marionette}}) demonstrate that these bitflips can be used to successfully mount security attacks that take over a computing system, steal secret data one does not have access to, or corrupt important data to render an application (e.g., a safety-critical ML/AI workload) useless or dangerous.

\begin{figure}[!ht]
    \centering
    \includegraphics[width=\linewidth]{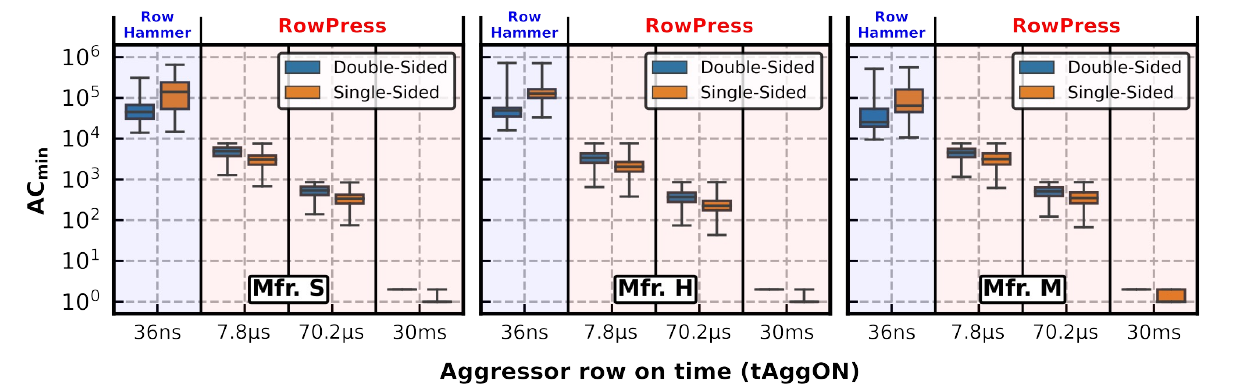}
    \caption{\textbf{Distribution of the number of activations required to induce a bitflip ($AC_{min}$) with RowHammer and three representative cases of RowPress at $80^{\circ}C$ across 164 DDR4 DRAM chips.} Adapted from~\cite{luo2023rowpress}.}
    \label{fig:rowhammer-rowpress}
\end{figure}

Unfortunately, RowHammer is {\em not} the only known prominent read disturbance phenomenon in DRAM {\em anymore}. We recently demonstrated, in an ISCA 2023 paper~\cite{luo2023rowpress,luo2024rowpress}, that modern DRAM chips are vulnerable to {\em RowPress}, a phenomenon where keeping a DRAM row active (i.e., open) induces bitflips in physically nearby rows. 
RowPress greatly amplifies read disturbance, reducing the number of activations required to induce a bitflip by one-two orders of magnitude (Fig.~\ref{fig:rowhammer-rowpress}), and enabling the inducing of bitflips in real systems even when DRAM chips are protected against RowHammer~\cite{luo2023rowpress}. Inspired by our demonstration of RowPress, recent device-level works aim to understand and model the underlying causes of the RowPress phenomenon~\cite{Zhou2024Unveiling,Zhou2024Understanding}.

\begin{figure}[!ht]
    \centering
    \includegraphics[width=\linewidth]{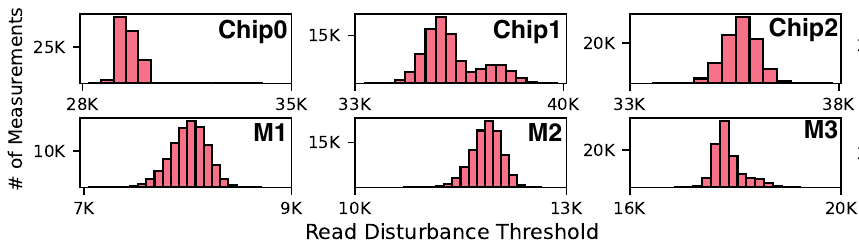}
    \caption{\textbf{Read disturbance threshold of a row in each tested HBM2 chip (Chip0-2) and DDR4 module (M1-3) over 100,000 repeated measurements.} Adapted from~\cite{olgun2025variable}.}
    \label{fig:vrd}
\end{figure}

Very recently, at HPCA 2025~\cite{olgun2025variable}, we experimentally demonstrated on 164 real COTS DDR4 and HBM2 DRAM chips a new phenomenon, {\em Variable Read Disturbance (VRD)}, that makes handling DRAM read disturbance harder: read disturbance vulnerability (number of activations required to induce a bitflip) of a DRAM row changes dynamically and unpredictably, as Fig.~\ref{fig:vrd} shows. We find that the read disturbance vulnerability of a row can vary by 3.5$\times{}$\iey{,} and the worst-case vulnerability of a row can take 94,467 measurements \iey{t}o determine. This, in turn, makes it hard to determine a safe threshold of number of activations at which a protection mechanism should kick in. At a high level, VRD is similar to VRT (variable retention time)~\cite{liu2013experimental,yaney1987meta,restle1992dram}, which leads to unpredictable dynamic changes in data retention times of DRAM cells, causing difficulties in determining safe refresh rates. VRT required the addition of ECC into DRAM chips, and our analysis suggests that properly handling VRD will require more complexity and guardbands in DRAM chips~\cite{olgun2025variable}. A device-level understanding of VRD is yet to be developed.

Clearly, DRAM technology scaling is getting much worse. How do we solve the scaling problems and maintain robust operation without losing performance or energy? The answer is not easy, especially since new failure mechanisms are likely to get discovered and could affect chips already in the field. Industry introduced various solutions to tackle RowHammer over the past decade, and the solutions have become increasingly complex (and likely more robust). Various implementations of PARA~\cite{kim2014flipping} and TRR~\cite{frigo2020trrespass,hassan2021utrr,micron2016trr} were initially employed in DDR3/DDR4 memory controllers and DRAM chips, and shown to be insecure~\cite{frigo2020trrespass,hassan2021utrr}. RFM~\cite{jedec2020jesd795} was introduced for DDR4, making the memory controller more complex. More recently, in the DDR5 standard (April 2024), JEDEC adopted {\em PRAC (per row activation counters)}~\cite{canpolat2025chronus,canpolat2024understanding,kim2023ddr5,jedec2024jesd795c}, a solution framework where each DRAM row has an associated activation counter stored in the DRAM chip. The DRAM chip internally implements a controller that 1) increments the activation counter of each row and 2) takes preventive actions to avoid bitflips, if the counter is above a threshold. Our recent works~\cite{canpolat2025chronus,canpolat2024understanding} analyze the security \& performance of PRAC, showing that its overheads can be large because DRAM chips cannot {\em autonomously} perform RowHammer maintenance and mitigation operations with low overheads.

Industry's PRAC solution, despite all its overheads \& downsides, is a move towards a more memory-centric {\em system-memory co-design} approach to handling DRAM technology scaling issues, as we had argued for at IMW 2013~\cite{mutlu2013memory}. This is because PRAC incorporates a slightly intelligent controller inside the DRAM chip that understands characteristics of DRAM cells and tries to ensure robust operation. Unfortunately, we believe currently implemented solutions are {\em not} memory-centric {\em enough}. A DRAM chip/system has no mechanism today to {\em completely autonomously} perform maintenance \& optimization operations (e.g., RowHammer/RowPress/VRD mitigation, intelligent refresh, memory scrubbing, profiling of memory cells for errors) internally, without requiring support from the memory controller (MC). MC dictates when a DRAM chip should perform refresh or RowHammer mitigation (unless the chip signals an error with a heavy-handed ALERT\_n pin~\cite{canpolat2025chronus,canpolat2024understanding}, which blocks the entire chip from being accessed). To enable efficient and flexible solutions to be implemented autonomously in DRAM, we need a better DRAM interface.

\begin{figure}[H]
    \centering
    \includegraphics[width=\linewidth]{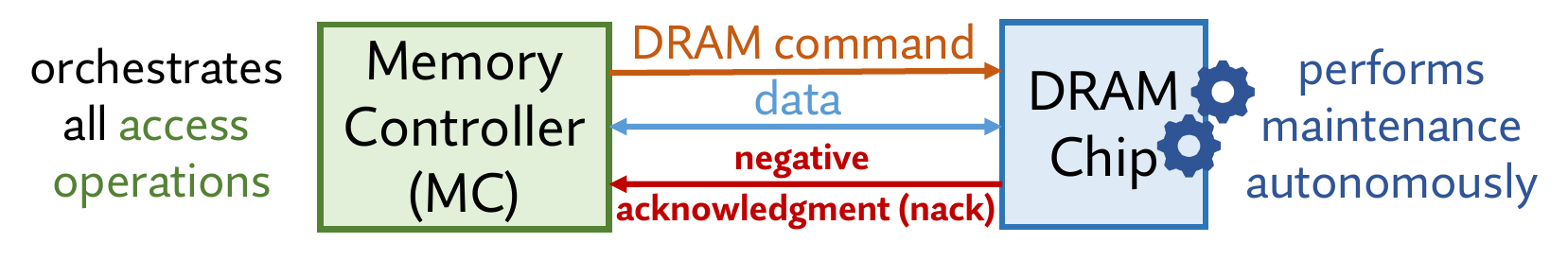}
    \caption{\textbf{Overview of Self-Managing DRAM (SMD).} Adapted from~\cite{olgun2024selfpresentation}.}
    \label{fig:smd}
\end{figure}

Our recent work at MICRO 2024~\cite{hassan2024self}, {\em Self Managing DRAM (SMD)}, introduces a more memory-centric interface and architecture that enables autonomous in-DRAM maintenance operations by transferring the responsibility of controlling maintenance operations from the memory controller to the SMD chip. To enable this, we make a single, simple modification to the DRAM interface (Fig.~\ref{fig:smd}), such that an SMD chip rejects MC requests to DRAM regions (e.g., a subarray or a bank) under maintenance, while allowing memory accesses to other DRAM regions. Thus, SMD enables 1) implementing new in-DRAM maintenance mechanisms (or modifying existing ones) with no further changes in the DRAM interface, MC, or other system components, and 2) overlapping the latency of a maintenance operation in one DRAM region with the latency of accessing data in another. Our results show that SMD provides large performance and energy benefits (when used to optimize refresh, RowHammer mitigation, \iey{and} memory scrubbing) while also improving system robustness across many workloads. Importantly, SMD enables easier adoption of innovative ideas to manage DRAM: a manufacturer can implement optimized mechanisms completely inside the DRAM chip without requiring changes to the DRAM interface or the MC. We believe that SMD can enable practical adoption of future innovative ideas in DRAM design and inspire better, more memory-centric, ways of partitioning work between memory and processor chips.

\section{System and Application Scaling}

With growing dataset sizes and computation needs of modern and emerging applications, systems designed to execute such applications need to scale cost-effectively to accommodate the memory \& computation demands. Unfortunately, scaling the systems (and hence applications) to much larger sizes is expensive and wasteful today in terms of energy, cost, and hardware real estate. The main culprit is the dichotomy between processing and memory: ideally we would like to add more memory capacity \& bandwidth as we add more computation capability~\cite{ahn2015scalable, ahn2023retrospective}, but doing so is expensive and wasteful due to the large separation and thin connectivity between computation and memory units today, caused by the processor-centric paradigm. As we add more memory and computation {\em separately}, we need to support higher bandwidth between them, which comes at 1) large monetary cost due to increased pin counts and 2) large energy and performance costs due to large amounts of data movement (Section~\ref{sec:problem}). With a memory-centric design, computation is placed inside memory chips (e.g., in 3D-stacked memory) and, thus, both memory bandwidth and computation capability can be proportionally increased to more efficiently scale up a system~\cite{ahn2015scalable, ahn2023retrospective, he2025papi, gu2025pim}. 

A major reason why processor-centric systems need high cost and high power consumption to scale up is because many key applications today are very data-intensive, in a way that renders much of the cache hierarchy ineffective (and thus adding more processors is very wasteful since most of a processor chip consists of caches and more memory bandwidth is required to support high memory demands). For example, major kernels in generative AI~\cite{he2025papi,gu2025pim, zhou2022transpim, park2024attacc,seo2024ianus,yun2024duplex,heo2024neupims,brown2020language,devlin2019bert,goodfellow2014generative}, graph analytics~\cite{ahn2015scalable,PEI,nai2017graphpim,besta2021sisa,salihoglu2013gps,tian2013from,low2012distributed}, and data analytics~\cite{seshadri2017ambit,deoliviera2021damov,gomez2022benchmarking,boroumand2022polynesia} workloads have low arithmetic intensity (i.e., low number of operations performed for each byte fetched from memory) due to the sparse and irregular nature of memory accesses and relatively small amount of computation needed.

Enabling computation capability in memory (i.e, {\em processing in memory}, or {\em PIM}) can overcome these challenges and enable efficient system and application scaling by improving many important metrics {\em at the same time}, including energy, performance, system-level hardware area efficiency, security, and even sustainability (by potentially eliminating large amounts of hardware wasted on processor-centric latency tolerance structures). There are two PIM types~\cite{mutlu2020modern,mutlu2019processing}: processing {\em near} memory (PNM) and processing {\em using} memory (PUM). PNM adds computational logic close to memory structures (e.g., in a DRAM chip, next to each bank, or at the logic layer of 3D-stacked memory~\cite{HMC2, HBM, lee2016simultaneous, hbm2, hbm3, kim2024present, ahn2015scalable,PEI,hsieh2016transparent,boroumand2018google,boroumand2022polynesia,boroumand2021google,boroumand2019conda,singh2021fpga,oliveira2022accelerating}). PUM performs computation by exploiting the analog operational properties of the memory circuitry. We believe both approaches are important to explore and enable as they have different tradeoffs: PNM can enable a wider set of functions (including complete processors) to be more easily implemented and exploited near memory due to its use of conventional logic, whereas PUM 1) more fundamentally reduces data movement by performing computation {\em inside} the memory arrays and 2) can fully exploit the large internal bandwidth and bit-level parallelism available \emph{inside} the memory arrays. We give examples of both PNM and PUM approaches, with a focus on DRAM, but refer the reader to our detailed overview paper for more information~\cite{mutlu2020modern}.

\subsection{Processing Near DRAM}

There are two major approaches to PNM inside a DRAM chip: adding logic near memory arrays in planar DRAM or in the logic layer of a 3D-stacked DRAM technology. Some commercially available PNM architectures, e.g. the UPMEM PIM system~\cite{gomezluna2021benchmarking,gomez2021benchmarkingcut,gomez2022benchmarking,gomez2023evaluating,gomez2022machine} take the planar DRAM approach, which has the advantage of providing logic in very {\em high-capacity} memory chips. However, it is difficult to fabricate low-cost, high-performance, and energy-efficient logic in DRAM fabrication process and the existing DRAM process does not enable enough metal layers to perform good communication across different PNM units or complex operations that require multiple metal layers. Yet, we believe such architectures are still critical to investigate and enable, with a focus on overcoming especially communication and architectural limitations. Several recent works demonstrate promising results with even the unoptimized first generation UPMEM PIM system\redt{~\cite{gomezluna2021benchmarking,gomez2021benchmarkingcut,gomez2022benchmarking,gomez2023evaluating,gomez2022machine, giannoula2024pygim,rhyner2024analysis,gogineni2024swiftrl,gupta2023evaluating,diab2023framework,frouzakis2025pimdal,lee2024spid,lee2024analysis,giannoula2022sparsep,oliveira2022accelerating,lim2023design,giannoula2022towards,diab2022high,chen2023uppipe,lavenier2020variant,lavenier2016blast,kang2023pim,baumstark2023accelerating,baumstark2023adaptive,bernhardt2023pimdb,nider2022bulk,das2022implementation,zarif2023offloading}} and some works develop methods to make such a system even better~\cite{lee2024pim,son2025pimnet,hyun2024pathfinding}. The availability of real PNM hardware also has enabled researchers to develop programming frameworks, compilers, and libraries for PNM systems~\cite{chen2023simplepim,oliveira2023transpimlib,khan2024cinm,oliveira2023dappa,noh2024pid}. As such, real PNM hardware acts as a catalyst for changing the computing paradigm. 

\begin{figure}[hb]
    \centering
    \includegraphics[width=\linewidth]{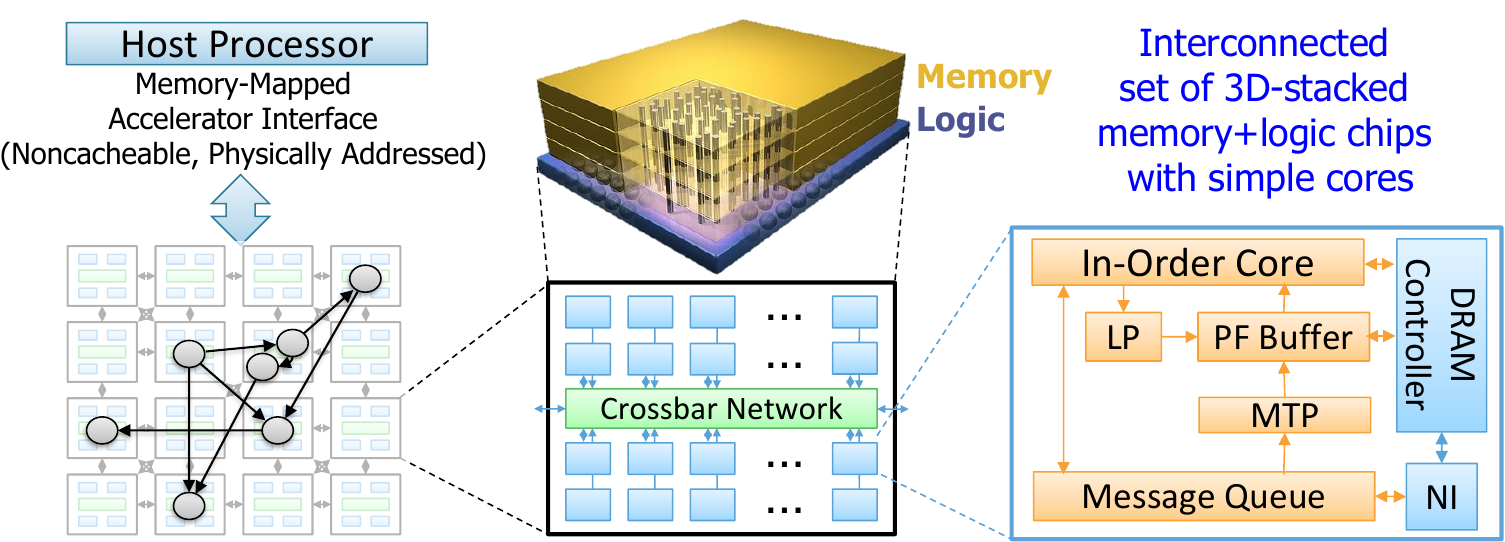}
    \caption{\textbf{Overview of the Tesseract system for graph processing.} Adapted from~\cite{mutlu2023-mcc-dac-talk}.}
    \label{fig:tesseract_arch}
\end{figure}

Increasingly maturing 3D-stacked integration/packaging technologies can enable more efficient PNM because 1) high-quality memory layers can be stacked on top of a high-quality logic layer, fabricated using a high-quality logic process, similar to a high-performance microprocessor, and 2) vertical connections between memory and logic layers can be smaller, more abundant and robust. Hybrid bonding\redt{~\cite{fujun2020stacked,niu2022184qps,schmidt1998wafer,kagawa2016novel}} and monolithic 3D technologies~\cite{ghiasi2022revamp3d, aly2015energy} are two promising advanced packaging/integration technologies that can enable efficient PNM. Prior works demonstrate large performance and energy benefits from using the logic layer to execute major data-intensive applications, like graph analytics~\cite{ahn2015scalable,PEI,nai2017graphpim,besta2021sisa}, or functions/kernels/layers in many different workloads, including genome analysis~\cite{kim2018grim,cali2020genasm,cali2022segram,diab2023framework}, machine learning~\cite{brown2020language,devlin2019bert,boroumand2021google,oliveira2022accelerating,gomez2023evaluating,rhyner2024analysis,gogineni2024swiftrl,heo2024neupims,zhou2022transpim,park2024attacc,seo2024ianus,yun2024duplex}, video processing~\cite{challapalle2020x,boroumand2018google,boroumand2021google}, compression/decompression~\cite{boroumand2018google,boroumand2021google}, database analytics~\cite{boroumand2019conda,amiraliphd,boroumand2016lazypim,boroumand2022polynesia,drumond2017mondrian,hsieh2016accelerating,RVU}. Figure~\ref{fig:tesseract_arch} demonstrates a high-level view of the Tesseract system we proposed in ISCA 2015~\cite{ahn2015scalable}, an accelerator that is comprised of a distributed system of 3D-stacked memories, which, in a coordinated manner, perform graph analytics computations by minimizing data movement and enabling performance that is proportional to memory capacity and bandwidth, as both scale linearly with more PNM compute units in the logic layer~\cite{ahn2015scalable,ahn2023retrospective}. Tesseract improves graph analytics performance by 13.8X while also reducing energy consumption by more than 8X, compared to a powerful state-of-the-art processor-centric system~\cite{ahn2015scalable}. Today, due to advances in packaging/integration technologies, the envisioned Tesseract system is much closer to being real (and can benefit many workloads). Similarly, systems envisioned for accelerating mobile workloads~\cite{boroumand2018google} and neural network execution~\cite{boroumand2021google} by offloading data-intensive functions/layers to 3D-stacked memories provide large performance and energy benefits compared to the best processor-centric systems. 

\begin{figure}[!h]
    \centering
    \includegraphics[width=\linewidth]{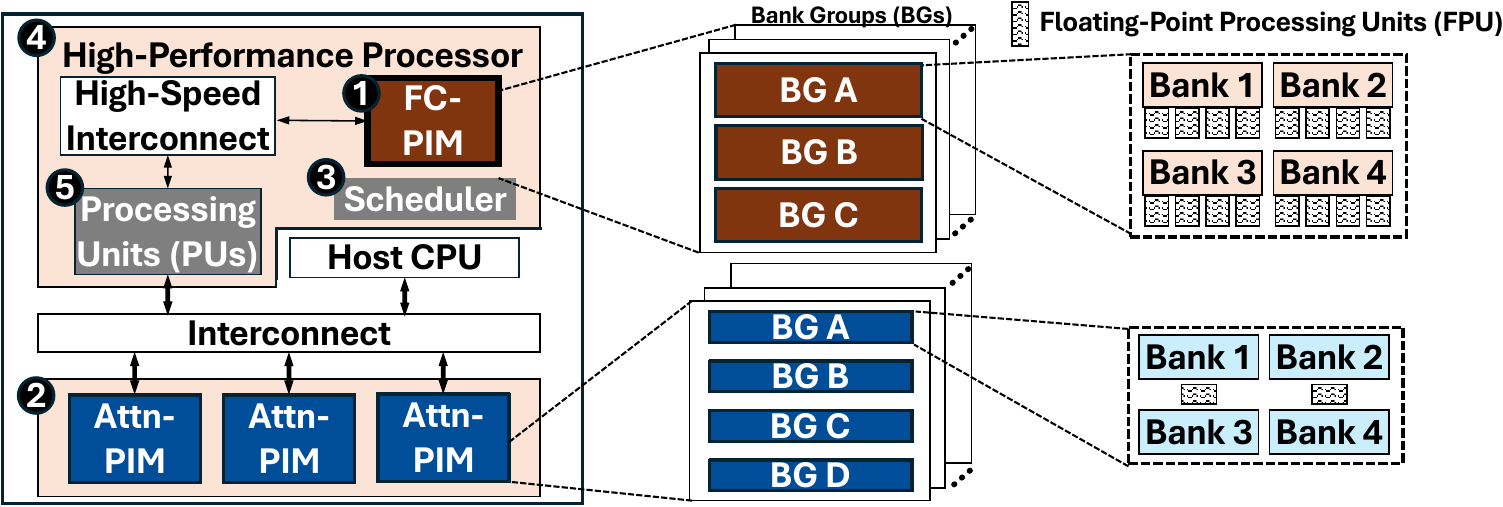}
    \caption{\textbf{Overview of the PAPI LLM Inference System.}
    Adapted from~\cite{he2025papi}.\vspace{-6pt} 
     } 
    \label{fig:papi_arch}
\end{figure}

Recently, several works demonstrated large performance, energy, and cost improvements from using PNM DRAM chips (that employ near-bank accelerators) to design a system for executing large language model (LLM) inference tasks. We discuss two major works published at ASPLOS 2025, PAPI~\cite{he2025papi} and CENT~\cite{gu2025pim}. The PAPI system (Fig.~\ref{fig:papi_arch}) uses two types of PNM DRAM chips, catering to two different kernel types in LLM inference workloads that perform speculative decoding: FC-PIM units \dingOne{} handle memory-bound fully-connected (FC) kernels that have high computational demands and are designed to have more computational capability inside the DRAM chip at the expense of some capacity. Attn-PIM units \dingTwo{} handle memory-bound attention kernels and store the large KV cache: they are designed to have less computational power but provide very large memory capacity. The PAPI scheduler \dingThree{} inside the high-performance processor \dingFour{} dynamically identifies which kernel should be executed in which PIM unit or the high-performance processing units (PUs) \dingFive{}, e.g., a GPU, based on the type of kernel and its arithmetic intensity, which varies at runtime. PAPI is a scalable system as 1) both the memory capacity/bandwidth and computational power of the FC-PIM and Attn-PIM units can be increased proportionally, and 2) Attn-PIM units  enable extra large capacity as they are disaggregated from the rest of the system. PAPI provides large performance and energy benefits over the best prior LLM inference systems~\cite{he2025papi}, with its careful use of multiple different types of PNM units along with powerful processor-centric units (e.g., GPUs or TPUs).

\begin{figure}[!ht]
    \centering
    \includegraphics[width=\linewidth]{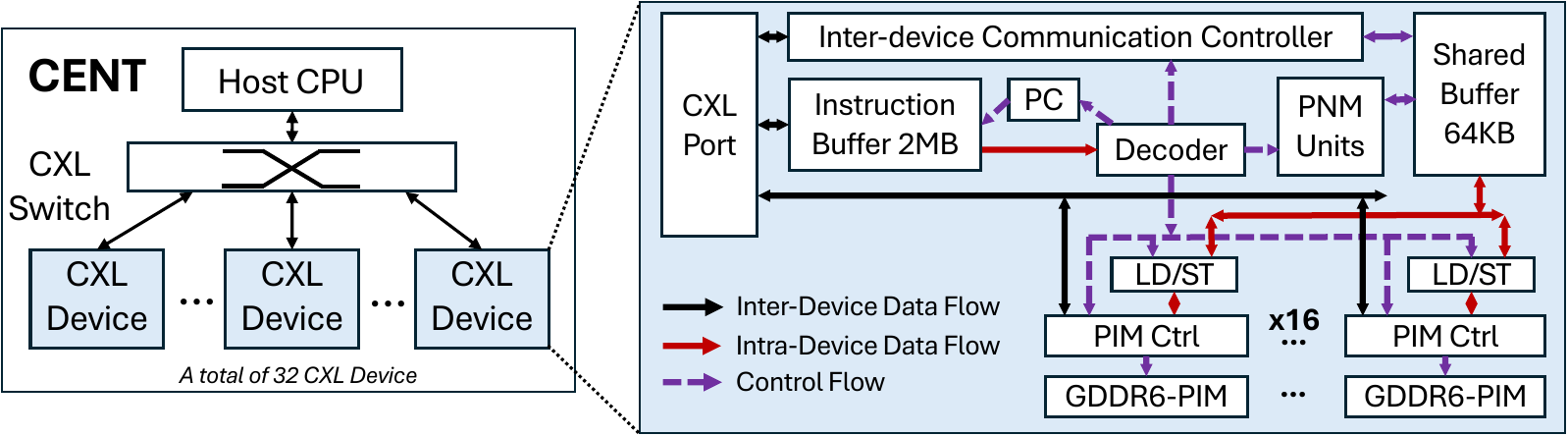}
    \caption{\textbf{Overview of the CENT LLM Inference System.} Host CPU drives 32 CXL devices, each having a CXL controller, PNM units, \iey{and} 16 GDDR6-PIM chips. The LLM inference task is partitioned between PNM units and GDDR6-PIM chips. CENT provides communication mechanisms within and across CXL devices to coordinate and scale computation. Adapted from~\cite{gu2025pim}.}
    \label{fig:cent_arch}
\end{figure}

The CENT system (Fig.~\ref{fig:cent_arch}) provides similar memory capacity, bandwidth, and computation scalability benefits by disaggregating a large number of PNM units (some in GDDR6-PIM chips \& some in CXL~\cite{van2019hoti} controllers) and providing communication primitives using a flexible interface (CXL) to enable communication between PNM units. The LLM inference task is distributed across the many CXL devices, in a manner similar to Tesseract that distributes graph computations across 3D-stacked PNM units. Computational tasks that require high memory bandwidth are executed inside the accelerators in GDDR6-PIM chips; tasks that require aggregation or expensive operations are executed inside the PNM units in CXL controllers. CENT {\em eliminates} the need for expensive GPUs by enabling a large number of high-capacity and high-computational-power PNM-enabled CXL devices to perform LLM inference in a coordinated manner, improving throughput by 2.3X, hardware cost by 2.4X, and tokens per dollar by 5.2X over a state-of-the-art system that uses GPUs~\cite{gu2025pim}.

\vspace{-7pt}
\subsection{Processing Using DRAM}
\label{sec:pud}
\vspace{-2pt}
Processing Using DRAM (PUD) systems use the operational principles of DRAM to perform primitive operations (e.g., data copy, initialization, bitwise operations), on top of which different applications and software stacks can be built. Early works~\cite{seshadri2013rowclone,seshadri2015fast,seshadri2017ambit} introduced PUD using first principles and circuit \& architectural simulations. RowClone~\cite{seshadri2013rowclone} demonstrates that consecutively activating two rows in the same DRAM subarray in quick succession performs copying of one row's content into the other. 
Ambit~\cite{seshadri2015fast,seshadri2017ambit,seshadri2019dram,seshadri2016buddy} demonstrates that 
1)~concurrently activating three DRAM rows leads to the computation of the bitwise MAJority function (and thus AND and OR) on the contents of the three rows
and
2)~bitwise NOT of a row can be performed through the sense amplifier, with modifications to DRAM circuitry. Ambit provides a DRAM chip architecture that can exploit such triple-row activation (TRA), NOT, and RowClone operations. SIMDRAM~\cite{hajinazar2021simdram} shows that, via a new software/hardware cooperative framework, \emph{any} 
operation (e.g., multiplication, division, convolution) that can be expressed as a logic circuit consisting of AND, OR, NOT gates can be implemented and seamlessly programmed using the Ambit substrate. MIMDRAM~\cite{deoliveira2024mimdram} makes the Ambit substrate much more flexible and easier to exploit, by enabling finer-granularity operations than the full row (with changes to DRAM architecture~\cite{olgun2024sectored}) and providing compiler support that transparently transforms applications to exploit bulk-bitwise execution in DRAM.

\begin{figure}[!ht]
    \centering
    \includegraphics[width=\linewidth]{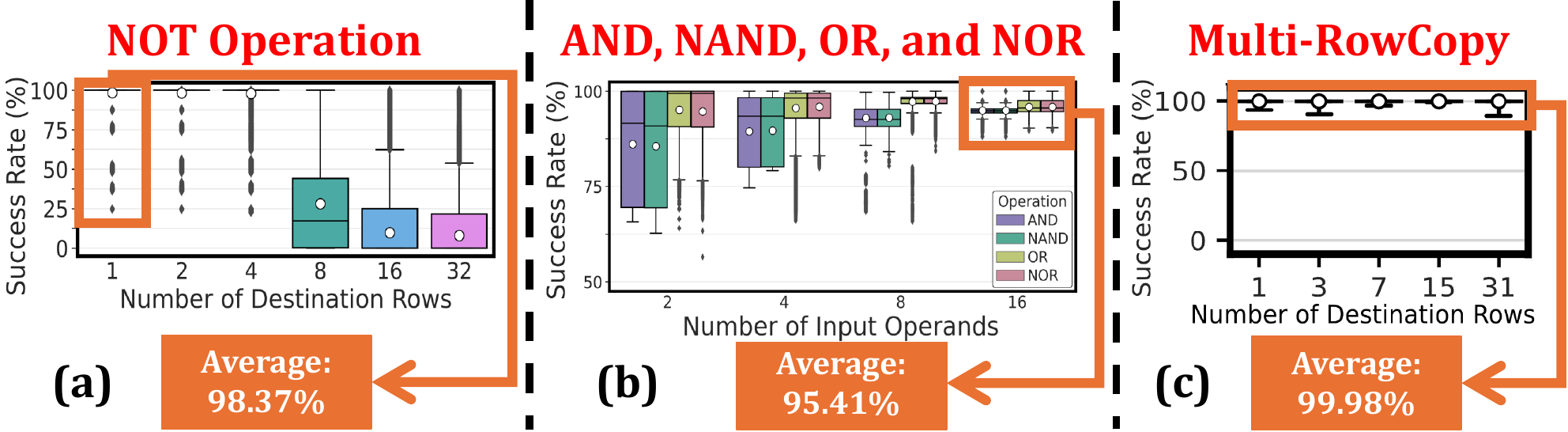}
    \caption{\textbf{Success rates of various operations in COTS DRAM chips: (a) NOT with varying destination rows, (b) AND, NAND, OR, NOR with varying input operands, (c) Multi-RowCopy with varying destination rows, as measured in 224, 224, and 120 COTS DRAM chips, respectively.} Adapted from~\cite{mutlu2024memory}. More info in~\cite{yuksel2024functionally,yuksel2024simultaneous}.}
    \label{fig:fcdram_simra_results}
\end{figure}

Fascinatingly, operations envisioned by these PUD works can \emph{already} be performed in \emph{real unmodified} \iey{COTS} DRAM chips.~Multiple recent works~\cite{yuksel2024functionally,yuksel2024simultaneous,yuksel2023pulsar} experimentally demonstrate previously-unknown capabilities in COTS DRAM chips. These capabilities arise from the operational principles of DRAM circuitry that are exercised by violating the manufacturer-recommended timing parameters via a flexible memory controller~\cite{olgun2023dram, hassan2017softmc,gao2019computedram,olgun2022pidram,gao2022frac}.
In particular, one can simultaneously activate {\em many} DRAM rows in state-of-the-art DRAM chips due to the hierarchical design of the row decoder circuitry~\cite{yuksel2024simultaneous,yuksel2023pulsar,bai2022low,weste2015cmos,turi2008high}.
Exploiting such simultaneous row activation, we~\cite{yuksel2024functionally,yuksel2024simultaneous,yuksel2023pulsar}
demonstrate that COTS DRAM chips are capable of 1) performing functionally-complete bulk-bitwise Boolean operations: NOT, NAND, and NOR, 2) executing up to 16-input AND, NAND, OR, and NOR operations, and 3) copying the contents of a DRAM row (concurrently) into up to 31 other DRAM rows. We evaluate the robustness of these operations across data patterns, temperature, and voltage levels. Our results (Fig~\ref{fig:fcdram_simra_results}) show that COTS DRAM chips can perform these operations at high success rates ($>$94\%) and data copy almost perfectly ($>$99.98\% success rate).
These fascinating findings demonstrate the fundamental computation capability of real DRAM chips, even when they are {\em not} designed for this purpose, and provide a solid foundation for building new and robust PUD mechanisms into future DRAM chips and standards.

\vspace{-5pt}

\redt{PUD can be used to generate true random numbers (TRNs) at low hardware cost, high throughput, and low energy~\cite{olgun2021quac,bostanci2022dr,kim2019d,talukder2019exploiting,tehranipoor2016robust}. For example, QUAC-TRNG~\cite{olgun2021quac} demonstrates that s}imultaneous activation of multiple rows in DRAM can be used for generating true random numbers at high throughput (e.g., 3.44 Gb/s per DRAM channel~\cite{olgun2021quac}),
widening the workloads supported by PIM systems (e.g., security-critical workloads) and enabling secure execution support for PUD systems
that do \emph{not} necessarily have dedicated TRN generation (TRNG) hardware.
Best prior TRNG using COTS DRAM chips generates TRNs by simultaneously activating four rows~\cite{olgun2021quac}. Our ongoing work~\cite{mutlu2024memory} experimentally studies the simultaneous activation of 2, 8, 16, and 32 rows in a subarray in COTS DRAM chips, showing that 8- and 16-row activation-based TRNG designs provide 1.25$\times$ and 1.06$\times$ higher throughput than the state-of-the-art.
\setstretch{0.968}
\section{Enabling Adoption}

New hardware is never easy to adopt, if it requires changes in software.
In memory-centric computing (MCC), we are not only introducing new hardware, but a new, different paradigm that performs computation in places never before considered by software (or hardware). As such, the biggest adoption issues of \iey{MCC} systems are related to software and the interfaces between software \& hardware and system components. We, therefore, believe it is critical to focus on designing frameworks for enabling MCC, including \iey{programming} frameworks, new workloads and algorithms, compilers, system software, runtime systems, and evaluation prototypes. Such frameworks can enable easy use, programming, and exploitation of PIM. To this end, we are developing new programming frameworks~\cite{chen2023simplepim,oliveira2023dappa,zhou2022transpim,oliveira2023transpimlib}, compilers~\cite{deoliveira2024mimdram,hsieh2016transparent,oliveira2025proteus}, system-level mechanisms~\cite{boroumand2016lazypim,boroumand2019conda,hsieh2016accelerating,hsieh2016transparent,ghiasi2022alp}, benchmarks~\cite{gomezluna2021benchmarking,gomez2021benchmarkingcut,gomez2022benchmarking,gomez2023evaluating,deoliviera2021damov}, and real evaluation prototypes~\cite{olgun2022pidram,olgun2023dram} for PIM systems, but there is still much more research and development that needs to be done, as described in our overview paper~\cite{mutlu2020modern}. 

We believe there is an evolutionary path to more easily adopt MCC. Instead of changing the paradigm overnight, we can incrementally introduce new operations and interfaces. For example, Self-Managing DRAM (Section~\ref{sec:memory_scaling})~\cite{hassan2024self} introduces a simple DRAM interface change with potentially very large long-term \& short-term benefits. Adopting it requires will and a shift into a more forward-looking mindset. Similarly, RowClone (Section~\ref{sec:pud})~\cite{seshadri2013rowclone} requires very small changes to DRAM chips and interface to be officially supported. Section~\ref{sec:pud} already showed that COTS DRAM chips can perform RowClone with almost perfect success rates even though they are {\em not} designed for this purpose. We believe adoption will become much easier once there are SMD chips or RowClone-capable chips, on top of which novel software and system mechanisms can be built. 

If {\em "insanity is doing the same thing over and over again and expecting different results"}~\cite{brown1983sudden,narcotics1981basic}, then we may have been insane since we have stuck to the processor-centric paradigm for so long at huge system performance, energy, area \& complexity costs. The good news is we seem to be a bit less insane today than a decade ago as we now have some compute-capable memories (e.g., PRAC, UPMEM, and DRAM PIM prototypes from various major companies), and packaging/integration technologies are on our side to make future systems more memory-centric. Do we have the will?

\section*{Acknowledgments}
This paper is an extended version of our invited paper and presentation in the ``DRAM'' focus session of the IMW 2025 conference~\cite{mutlu2025memory}.  
We thank the SAFARI Research Group members for providing a stimulating intellectual and scientific environment. We acknowledge the generous gifts from our industrial partners, including Google, Huawei, Intel, and Microsoft. This work, along with our broader work in Processing-in-Memory and memory systems, is supported in part by the Semiconductor Research Corporation (SRC), the ETH Future Computing Laboratory (EFCL), {ACCESS -- AI Chip Center for Emerging Smart Systems}, a Google Security and Privacy Research Award, and the Microsoft Swiss Joint Research Center.

\balance
\setstretch{0.956}
\bibliographystyle{IEEEtran}
\bibliography{refs}
\end{document}